\documentclass[%
 reprint,
superscriptaddress,
 amsmath,amssymb,
 aps,
]{revtex4-2}
\usepackage[hidelinks]{hyperref} 
\hypersetup{breaklinks=true}
\usepackage{graphicx}
\usepackage{dcolumn}
\usepackage{bm}

\begin{document}

\preprint{APS/123-QED}

\title{Avalanche Sensing via Kerr frequency comb in an Optical Microcavity}
\author{Chenchen Wang}
\email{cwang922@wisc.edu}
\affiliation{Department of Electrical and Computer Engineering, University of Wisconsin--Madison, Madison, Wisconsin 53706, USA}

\author{Qingyi Zhou}
\affiliation{Department of Electrical and Computer Engineering, University of Wisconsin--Madison, Madison, Wisconsin 53706, USA}

\author{Lan Yang}
\affiliation{Department of Electrical and Systems Engineering, Washington University, St. Louis, Missouri 63130, USA}

\author{Zongfu Yu}
\affiliation{Department of Electrical and Computer Engineering, University of Wisconsin--Madison, Madison, Wisconsin 53706, USA}


\date{\today}

\begin{abstract}
Sensors based on optical microcavities enhance light–matter interactions within an ultraconfined volume, enabling high-sensitivity detection across a wide range of sensing applications. In these systems, environmental perturbations modify the intrinsic resonance properties of the cavity, typically manifested as frequency shifts, linewidth broadening, or mode splitting. However, the minimum resolvable change in these spectral properties fundamentally limits the overall sensor sensitivity. Here, we propose a new avalanche sensing scheme enabled by Kerr nonlinearity. Instead of relying on the detection of frequency shifts, our approach exploits abrupt state transitions in a Kerr frequency comb to amplify weak perturbations. We provide a theoretical analysis of the underlying mechanism of this scheme and validate the concept through both coupled-mode theory (CMT) modeling and full-wave electromagnetic simulations.
\end{abstract}


\maketitle
\begin{figure*}[htbp]
    \centering
    \includegraphics[width=0.99\linewidth]{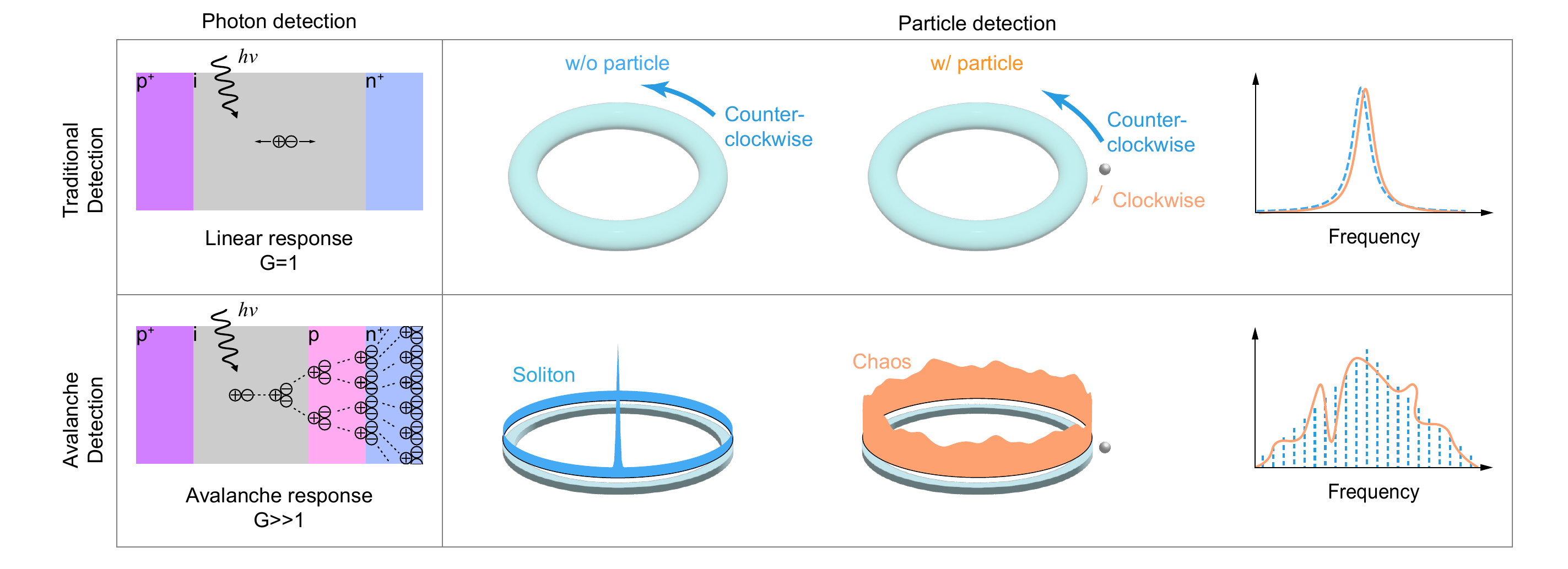}
    \caption{Principle of avalanche-inspired Kerr microcavity sensing.  Four-quadrant schematic comparing conventional and avalanche detection paradigms for photon/particle sensing. While traditional sensing detects a particle by resolving the minute perturbation it induces on the cavity resonance frequency, avalanche approach's sensitivity is enhanced by Kerr nonlinearity: the particle-induced shift of the cavity eigenfrequency drives the intracavity Kerr comb soliton across a state-transition boundary, converting an ultrasmall frequency perturbation into a macroscopic comb-state change.}
    \label{fig:1}
\end{figure*}
Optical microcavities with ultrahigh quality factors and tightly confined electromagnetic modes provide an effective platform for detecting extremely weak perturbations\cite{Vahala2003,nano15030206,Loyez2023,righini2016}. Such field enhancement facilitates precision measurements such as mechanical displacement sensing\cite{Gavartin2012}, nanometre-scale objects detection\cite{zhi2017}, and absorption spectroscopy at the single-particle level\cite{Heylman2016}. External disturbances in these systems typically perturb cavity resonance properties—most often as frequency shift\cite{vollmer2008},  linewidth broadenin\cite{shao2013}, or mode splitting\cite{Zhu2010}. The smallest resolvable change in these spectral features, however, imposes a fundamental limit on the sensitivity of conventional microcavity sensors\cite{Vollmer2012}. Although various strategies have been developed to amplify the sensing response—such as plasmonic enhancement of light–matter interactions \cite{Baaske2014}and sensitivity boosting near exceptional points\cite{Chen2017}—the underlying detection principle remains to compare slight resonance variations. Thus, the measurable signal continues to be constrained by the induced frequency perturbation level, causing practical detection precisions well below the theoretical limit\cite{Foreman:15}.

In parallel, Kerr optical frequency combs generated in high-$Q$ optical microresonators have emerged as a transformative platform that advances a broad spectrum of scientific and technological applications\cite{Fortier2019,Chang2022,Dalvit2024}. In these systems, a continuous‑wave (CW) pump laser drives a microresonator where Kerr nonlinearity balances group velocity dispersion. Governed by the Lugiato–Lefever equation (LLE), these systems exhibit rich nonlinear dynamics ranging from Turing patterns to spatiotemporal chaos. Under specific pumping and detuning conditions, these interactions stabilize into dissipative Kerr solitons (DKSs)—self-organized and highly coherent pulses circulating within the cavity\cite{Herr2013,Herr2014,Kippenberg2011}. Crucially, these dynamical states are not only hosted in the same high-$Q$ environment ideal for sensing but are also highly sensitive to the effective cavity resonance\cite{Pasquazi2018}. This unique combination suggests that the global dynamical state of a frequency comb may serve as a powerful probe for sensing applications.

Inspired by criticality-enhanced sensing and nonlinear dynamical amplification mechanisms\cite{Bruschini2019, chu2021,DiCandia2023,Needham2024}, here we introduce a nanoparticle sensing paradigm based on Kerr microcavities as shown in Fig.~\ref{fig:1}. In contrast to traditional sensing methods limited by resolving small perturbation-induced resonance shifts, our approach utilizes the macroscopic dynamical state of the frequency comb as the sensing observable. We demonstrate that subtle environmental perturbations, effectively acting as minute variations in cavity detuning, can destabilize the system's equilibrium and initiate a state transition driven by the accumulation over time. Analogous to an electronic avalanche, this mechanism enables the system to surpass the resolution limits of traditional frequency-domain measurements.


As a concrete implementation, we consider a label-free single-particle sensor based on a Kerr whispering-gallery-mode (WGM) resonator with anomalous dispersion, where the interplay between dispersion and Kerr nonlinearity supports stable soliton formation. The spatiotemporal evolution of the system is captured by the Lugiato–Lefever equation\cite{Godey2014} 
\begin{equation}
\partial_{\tau} \psi = -(1 + i\alpha) \psi - \frac{i}{2} \beta \partial_{\theta}^{2} \psi + i |\psi|^2 \psi + F \label{eq:1}
\end{equation}
where $\psi$ represents the complex envelope of the total intracavity field, $\tau$ represents dimensionless time, $\theta$ denotes the azimuthal angle of the ring resonator, $\alpha$ and $\beta$ are dimensionless frequency detuning and dispersion separately. The last term in this equation $F$ represents the dimensionless external pump field intensity. 

Under CW pumping, the steady-state solutions of Eq.~\ref{eq:1} can be obtained by setting all the derivatives to zero, thereby yielding\cite{Coen13}
\begin{equation}
    X = Y^{3} - 2\alpha Y^{2} + (\alpha^{2} + 1)Y,
    \label{eq:2}
\end{equation}
Here $X = |F|^{2}$ and $Y = |\psi|^{2}$ are the normalized pump and intracavity powers, respectively. The cubic steady-state relation in Eq.~\ref{eq:2} exhibits multistability admitting up to three real roots depending on the normalized detuning and the pumping power. Theoretical research demonstrates that the existence of DKSs is strictly confined to the multistable window yielding three real solutions. Specifically, these solutions correspond to a stable upper and lower branch, mediating an unconditionally unstable middle branch. Within this domain, the background field of DKSs corresponds to the lower equilibrium branch, whereas the localized pulse is associated with the upper branch\cite{Godey2014}. Therefore, when variations in $\alpha$ drive the system out of the multistability window with three real roots, DKSs vanish accordingly.

On the other hand, the adsorption of a nanoparticle onto the cavity surface is well described by first-order perturbation theory as a small shift in the resonance frequency, as expressed in Eq.~\ref{eq:3}\cite{Arnold03}
\begin{equation}
 \frac{\delta\omega}{\omega} 
\simeq 
\frac{-\alpha_{\mathrm{ex}} \left| \mathbf{E}_0(\mathbf{r}_{\mathrm{p}}) \right|^2}
{2 \displaystyle \int \epsilon(\mathbf{r}) \left| \mathbf{E}_0(\mathbf{r}) \right|^2 \mathrm{d}V } .
\label{eq:3}
\end{equation}
where $\delta\omega/\omega$ denotes the fractional resonance frequency shift induced 
by a particle located at ${\mathbf{r}}_{\text{p}}$, $\alpha_{\mathrm{ex}}$ is the excess polarizability of the particle, ${\mathbf{E}}_0({r})$ is the unperturbed cavity mode profile, and $\epsilon({\mathbf{r}})$ is the permittivity distribution of the unperturbed system. This mechanism has been extensively verified in conventional microcavity sensors that rely on frequency-shift–based detection. 

In our scheme, the Kerr soliton is biased near the critical edge of its existence range. Consequently, even a nanoparticle-induced slight variation in the normalized detuning $\alpha$ pushes the system across the bifurcation point, driving the intracavity field from the soliton state into a qualitatively distinct regime, such as chaos, Turing pattern or the trivial CW state. Fig.~\ref{fig:1} contrasts this avalanche response with conventional frequency-domain detection. The sensitivity gain stems not merely from the different observational domain, but from the cumulative nature of the nonlinear dynamics. While the immediate perturbation manifests as negligible deviations in both time and frequency domains, the subsequent nonlinear evolution amplifies this seed into a macroscopic state transition that is easily detectable.
\begin{figure}[htbp]
    \centering
    \includegraphics[width=0.99\linewidth]{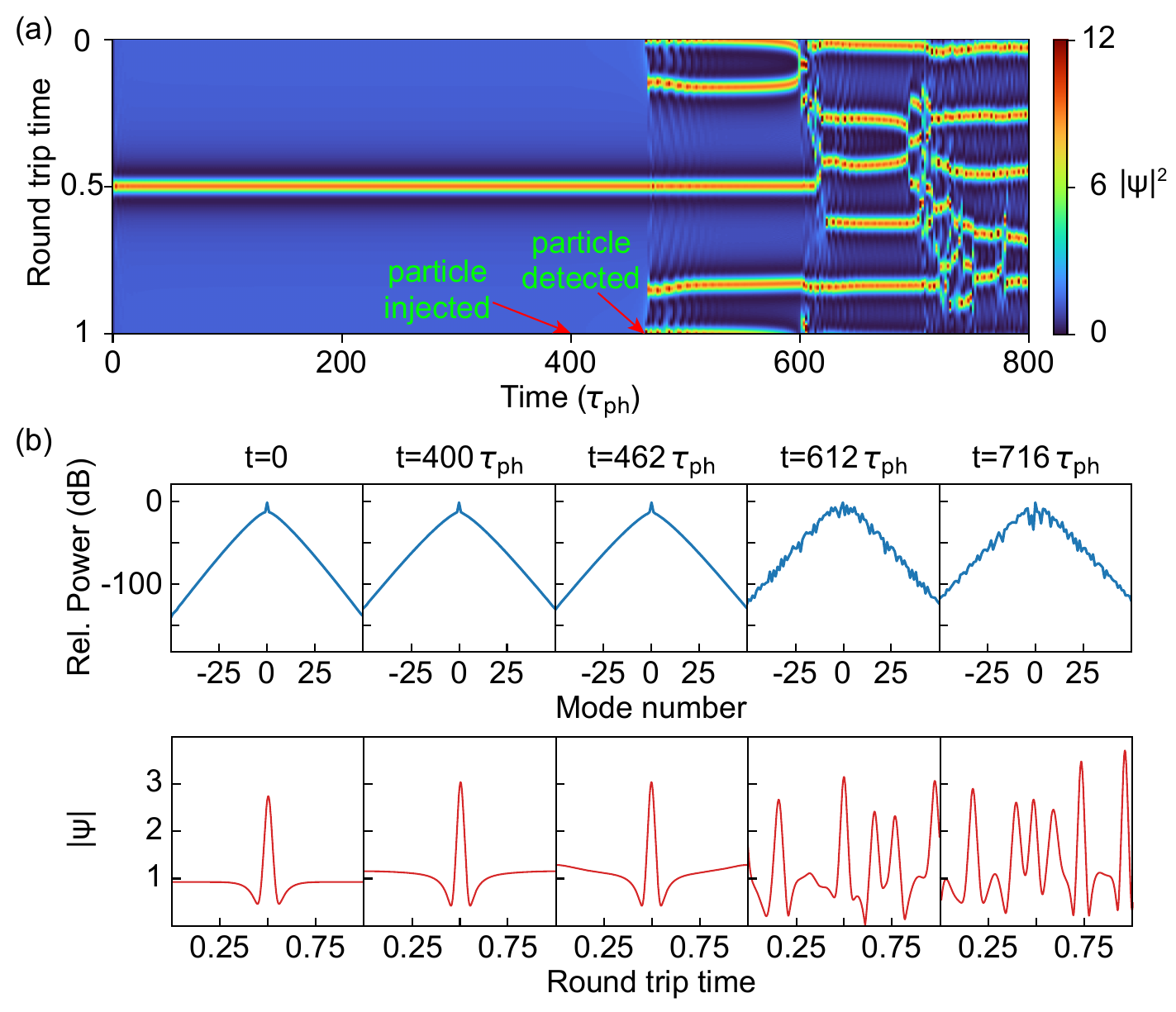}
    \caption{Numerical simulation of the sensing dynamics governed by the LLE. (a) Spatiotemporal evolution of the intracavity field intensity. (b) Selected spectral (top) and temporal (bottom) snapshots.}
    \label{fig:2}
\end{figure}

We validate the proposed sensing mechanism through two independent simulation approaches: coupled-mode theory (CMT) based modelling and full-wave electromagnetic simulations. We first present the results from the CMT based framework, which captures the normalized intracavity field evolution under the mean field and mode family approximations. The nanoparticle adsorption is treated as a step function shift in the modal resonance frequency. To initialize the system, we employ a soft stimulation strategy by seeding a short pulse. Since the sensing perturbation is introduced after the soliton reaches a steady state, this initialization yields a starting point effectively equivalent to that accessed via standard experimental detuning scans. As illustrated in Fig.~\ref{fig:2}(a), the system rapidly settles into a stable soliton that persists indefinitely under simulated noise conditions. At $t = 400\,\tau_{\mathrm{ph}}$ (photon lifetimes), the nanoparticle-induced frequency shift is introduced. Notably, the system exhibits a quasi-stable induction period: while the continuous-wave background shows minor fluctuations, the soliton pulse maintains a stable profile. However, after an integration time of approximately 100 $\tau_{\mathrm{ph}}$, the cumulative nonlinear evolution triggers a catastrophic destabilization. The localized pulse collapses, and the field rapidly evolves into a chaotic state. Fig.~\ref{fig:2}(b) presents spectral and temporal snapshots during this transition, illustrating the detailed pathway of the detection process.

\begin{figure}[htbp]
    \centering
    \includegraphics[width=0.99\linewidth]{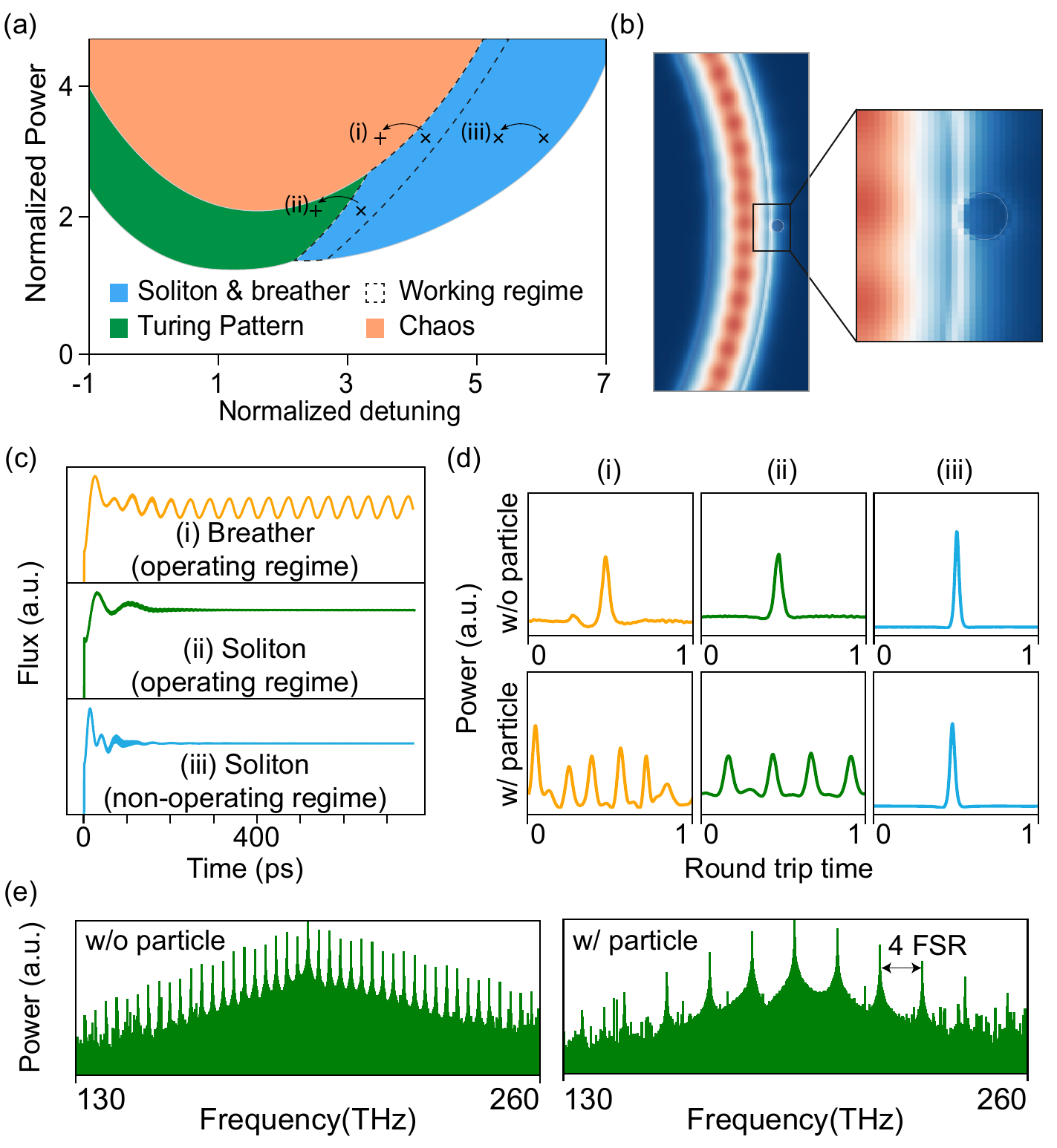}
    \caption{\label{fig:4}FDTD validation of the avalanche sensing mechanism. (a) Bifurcation diagram of Kerr-comb states, with the operating regime marked by a dashed box. Three sensing scenarios are annotated: (i)–(ii) correspond to proper biasing where the particle-induced shift drives the operating point across the boundary and triggers a comb-state transition, whereas (iii) illustrates a trivial mis-biasing case where the operating point moves but no transition occurs, rendering the particle event unobservable. (b) Electric-field distribution of the pumped mode with a zoomed-in view (right) highlighting the local perturbation from the nanoparticle. (c) Three distinct regimes observed in the simulation: breathing solitons (top) and two soliton regimes (middle and bottom), corresponding to scenarios (i), (ii), and (iii). (d) Intracavity states in these three regimes without (top) and with (bottom) a nanoparticle. (e) Spectral profiles in the (ii) regime: unperturbed (left) and perturbed (right).}
    \label{fig:3}
\end{figure}

The feasibility of the sensing mechanism is further verified by finite-difference time-domain (FDTD) simulations. To make the long-term nonlinear evolution computationally tractable, we employ a GPU-accelerated commercial FDTD solver (Tidy3D) and use the same soft stimulation strategy as in the CMT simulations to efficiently capture solitons. The simulation domain is also reduced to two dimensions, an approximation that has been shown to capture the essential physics of 3D Kerr comb dynamics accurately\cite{tidy3d}. 

Fig.~\ref{fig:3} presents results from FDTD simulations of a two-dimensional Kerr microring coupled to a straight waveguide. This framework considers the fundamental TE mode in the anomalous-dispersion regime over wavelengths from \(1.15\,\mu\mathrm{m}\) to \(2.3\,\mu\mathrm{m}\). We set the microring Kerr coefficient to $n_2 = 2\times10^{-20}~\mathrm{m}^2/\mathrm{W}$ and use the same linear refractive index, $n=1.5$, for both the microring and the nearby nanoparticle with radius $r=133~\mathrm{nm}$. Further simulation details are provided in the Supplemental Material.

In Fig.~\ref{fig:3}a, we first illustrate the bifurcation diagram of the Kerr comb states. As discussed above, the comb dynamics can be broadly categorized into chaotic states, Turing patterns (periodic signal of modulational instability), and solitons; when the soliton exhibits pronounced periodic amplitude modulation, we refer to it as a breather. Since a particle-induced effective red shift will decrease the normalized detuning, the preferred operating points reside in the soliton/breather region near the bifurcation boundaries (dashed boxes). Accordingly, we study three initial settings: (i) breathers adjacent to chaos, (ii) solitons adjacent to the Turing regime, and (iii) solitons initialized well away from the boundaries. Leveraging this framework, we first successfully reproduce the soliton and breather existence landscape. Fig.~\ref{fig:3}(c) illustrates three different generated regimes: a breather and two solitons. In the breather regime, the total intracavity flux is time-modulated. This global oscillation manifests locally as periodic variations in the pulse amplitude, whereas the solitons maintain constant flux and amplitude profiles. These results align well with previous theoretical predictions and experimental observations

Subsequently, as shown in Fig.~\ref{fig:3}(b), a high-index particle locally perturbs the pump field, inducing an effective red shift in the cavity resonance. To model the detection process within the FDTD framework, we perform parallel simulations under identical pumping and cavity conditions, differentiating only by the presence of the nanoparticle. Upon the particle perturbation, the three initial regimes (i)–(iii) exhibit distinct dynamical responses. As shown in Fig.~\ref{fig:3}(d), the two operating points inside the dashed boxes undergo clear state transitions: in case (i), the breather collapses into a chaotic state, whereas in case (ii), the soliton evolves into a Turing pattern regime. By contrast, in case (iii), when the operating point is outside the dashed boxes, the system remains in the soliton state on the particle perturbation, though with a slight amplitude variation. In our scheme, such a response corresponds to an indistinguishable particle event. These transitions are consistent with the bifurcation topology in Fig.~\ref{fig:3}(a). Fig.~\ref{fig:3}(e) contrasts the frequency spectrum in case (ii). While the unperturbed state (left) displays a dense single soliton comb, the perturbed state (right) reveals a Turing pattern spectrum with a 4-FSR mode spacing, perfectly consistent with the four-pulse spatial structure observed in Fig.~\ref{fig:3}(d).

\begin{figure}
\includegraphics[width=0.99\linewidth]{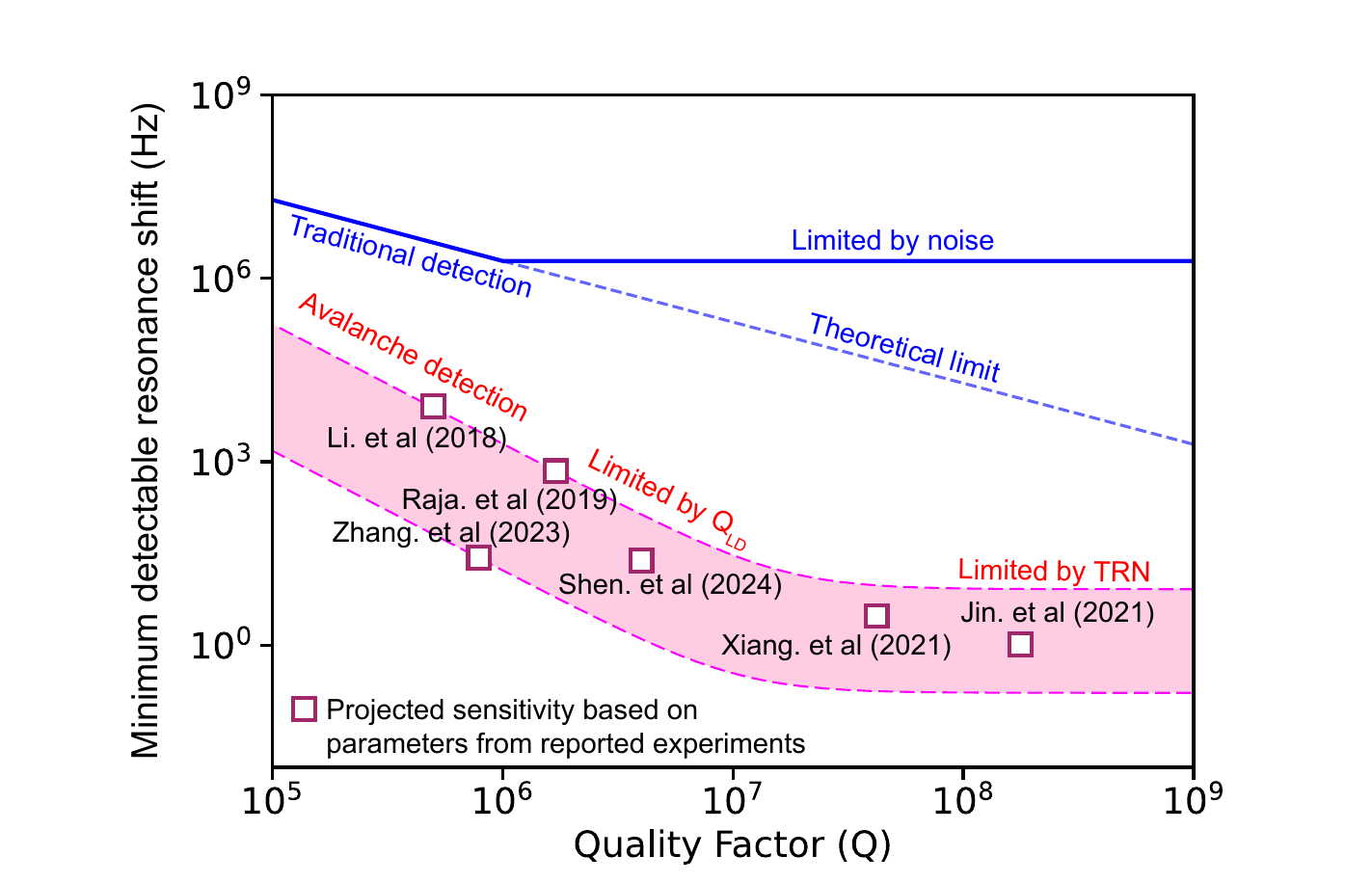}
    \caption{Sensitivity comparison between the traditional frequency shift approach and the proposed method. The blue line shows the limit of frequency shift detection. The pink region indicates the theoretical sensitivity range achievable with avalanche method. The purple open squares represent the projected sensitivities of avalanche detection if using experimental configurations reported in Refs\cite{Jin2021,Li2018,Raja2019,zhang2023,Shen24,Xiang2021}.}
    \label{fig:4}
\end{figure}

To evaluate the theoretical sensitivity, we compare our scheme against the conventional frequency shift method in Fig.~\ref{fig:4}. Despite its conceptual simplicity, the conventional frequency-shift method faces a fundamental barrier: when the particle-induced shift is smaller than $10^{-2}$ of the cavity linewidth\cite{shao2013}, the resulting spectral change becomes difficult to resolve. In practical experiments, the resolution is further limited by thermal noise, which—without specific optimization—typically restricts the achievable resolution to the megahertz regime\cite{Foreman:15}. These corresponds to the detection sensitivity limit of the frequency-shift method indicated by the blue dashed and solid lines in Fig.~\ref{fig:4}. In contrast, our approach monitors macroscopic changes of the intracavity state rather than relying on resolving small resonance shifts. The detection limit is instead determined by the stability of the detuning bias relative to the transition boundary. Recent progress in self-injection locking laser with microcavities has significantly narrowed the effective laser linewidth, thereby improving the precision of laser–cavity detuning control\cite{Kondratiev2023}. Based on our previous LLE-based and FDTD simulations, the full detection response to a particle event occurs within tens to hundreds of photon lifetimes, which is far shorter than the characteristic timescales associated with the technical-noise band of self-injection-locked lasers\cite{Voloshin2021}. At such ultrafast timescale of particle detection, laser noise reaches the white noise floor, corresponding to an effective fundamental linewidth\cite{DiDomenico10}. Here we use this fundamental linewidth to estimate the theoretical detection limit for our method. This linewidth scales as $Q^{-2}$, but at ultrahigh $Q$ it becomes increasingly dominated by thermo-refractive noise (TRN)\cite{Guo2022}. The pink band in Fig.~\ref{fig:4} illustrates this trend, with purple open squares markers denoting theoretical limits projected from parameters reported in recent experiments. Notably, these experiments either demonstrate frequency comb generation or operate in regimes compatible with comb formation. These results demonstrate a substantial sensitivity advantage over the conventional method.


In summary, we have proposed an avalanche sensing paradigm exploiting the nonlinear dynamics of dissipative Kerr solitons. Unlike linear methods constrained by spectral resolution, this mechanism leverages the intrinsic nonlinearity to temporally integrate infinitesimal. Our comprehensive validation, spanning independent CMT dynamics and full-wave FDTD simulations, confirms this mechanism's fidelity. Compatible with high-$Q$ microresonator platforms, this nonlinear method opens a new avenue for ultra-high-sensitivity metrology, pushing the boundaries of cavity-based sensing.

\bibliography{main}

\end{document}